\documentclass{article}
\newtheorem{theorem}{Theorem}
\newtheorem{conjecture}{Conjecture}

\begin{document}

\title{A Computer Search for \\ $N_{1L}$-Configurations}
\author{Martin Dowd \\ \texttt{MartDowd@aol.com}}
\date{}
\maketitle

\begin{abstract}
\boldmath
In an earlier paper the author defined $N_{1L}$ configurations, and stated
a conjecture concerning them which would lead to an improvement by a
constant factor to the sphere-packing bound for linear double error
correcting codes.  Here a computer search is presented, in an effort to
gather evidence on the conjecture.
\end{abstract}

\section{Introduction}

In \cite{Dowd07}, some configurations in binary linear codes, first
considered in \cite{Dowd88}, were further considered.  In particular, a
conjecture was made,
on the maximum number of rows of a configuration called an $N_{1L}$
configuration.  If true, the conjecture would yield an improvement by a
constant factor to the currently known bound (sphere packing bound) on the
maximum length $n$ of a double error correcting binary linear code of
redundancy $r$.

As in \cite{Dowd07},
the following conventions will be used.  Let ${\cal F}_q$
denote the finite field of order $q$, for a prime power $q$.  A binary code
of length $n$ is a subset of the vector space ${\cal F}_2^n$ over
${\cal F}_2$.  Such a code is linear if it is a subspace; if $k$ is the
dimension the redundancy $r$ is defined to be $n-k$.  Codewords are
generally denoted $v$, $w$, etc.  Positions are generally denoted $i$, $j$,
etc., with $1\leq i\leq n$.  As usual, $v_i$ denotes the element of
${\cal F}_2$ in position $i$ of $v$.  A vector in ${\cal F}_2^n$ will be
called a bit vector, of length $n$.
Generator matrices will be considered to be $k\times n$,
and parity check matrices $n\times r$.
A vector $v$ may be identified with its ``support'' $\{i:v_i=1\}$.
The Hamming weight $|v|$ of a vector
is the cardinality of the support.

An $N_1$ configuration is defined to be a set $S$ of weight 5 vectors,
of minimum distance 6, where there is a weight 5
``anchor'' vector $v$ and a position
$i\in v$, such that for $w\in S$, $i\in w$ and $|v\cap w|=2$.
An $N_1'$ configuration is a configuration of triples
(weight 3 vectors), which may be obtained from an $N_1$
configuration by deleting the positions of $v$.

Again as in \cite{Dowd07} a partial linear space is defined to be an incidence
matrix (matrix over ${\cal F}_2$), where two columns are incident to at most
one row.  Note that the requirement may equally be stated as, two rows are
incident to at most one column; the requirement is that no ``rectangle''
of 1's occur.  The following is observed in \cite{Dowd07}.

\begin{theorem}
An $N_1'$ configuration is a partial linear space,
of constant row weight 3, together with a partition of the rows into
4 or fewer parts, such that in each part the rows are disjoint.
\end{theorem}

Note that an $N_1$ configuration from which an $N_1'$ configuration
arises can be determined from the partition.  Clearly the maximum number of
rows in an $N_1$ configuration of length $n$ (equivalently an $N_1'$
configuration of length $n-5$) is $4\lfloor(n-5)/3\rfloor$.  This bound is
achieved in a 3-($n$,5,1) design; these exist for $n=4^m+1$ where $m\geq 1$
(see \cite{BJL}, theorem 6.9).

An $N_1$ configuration is said to be an $N_{1L}$ configuration if the linear
span of its rows and the anchor vector
has minimum weight 5, and an  $N_{1L}'$ configuration is an $N_1'$
configuration which arises from an $N_{1L}$ configuration.  Let $N_{1L}(n)$
denote the maximum number of rows in an $N_{1L}$ configuration in a linear
double error correcting code of length $n$.  The following conjecture was
made in \cite{Dowd07}.

\begin{conjecture}
$N_{1L}(n)$ is $\leq c_1(n-5)$ almost everywhere, for a
constant $c_1$ smaller than $4/3$.
\end{conjecture}

It was also shown that the conjecture yields an upper bound on the length of
a linear double error correcting code of redundancy $r$, better by a
constant factor than the sphere packing bound.  A computer search
showed that for $r\leq 8$, the conjecture holds with $c_1=2/3$.
However, an $N_{1L}(n)$ configuration with $n-5=60$ and 44 rows
was found in a cyclic code.

In this paper, a computer search is carried out for all $N_{1L}'$
configurations with $n-5<=18$ and $r<=14$, where $r$ is the number of rows.
This serves two purposes.  First, the existence of $N_{1L}$ configurations
for small $r,n$ pairs is determined.  Second, data is provided for possibly
inferring rules for an inductive proof of the conjecture.  It thus
represents an attempt to achieve a goal stated in \cite{Kaski}, that
``occasionally a practical algorithm and thereby a classification result
is obtained.''

We note here that there is a ``replication'' argument which shows that if a
ratio $r/(n-5)$ is achieved then it is achieved infinitely often.

\begin{theorem}
Let $M$ be an incidence matrix whose rows are divided into classes $M_i$,
$0\leq i\leq 3$.  For $m\geq 1$ let $M_i^m$ have $m$ copies of $M_i$ down
the diagonal.  Let $M^m$ be the matrix whose classes are the $M_i^m$.
\begin{enumerate}
\item[a.] if $M$ is an $N_1'$ configuration then $M^m$ is; and
\item[b.] if $M$ is an $N_{1L}'$ configuration then $M^m$ is.
\end{enumerate}
\end{theorem}

{\sl Proof:}
Part a follows readily using theorem 1 and is left to the reader.
For part b, theorem 5b of
\cite{Dowd07} will be used.  By a ``section'' of $M^m$
we mean the rows or columns of one of the replications.  Consider a sum of
rows of $M^m$.  If the weight of the sum is zero in a column section then
the corresponding rows can be deleted from the sum.  If there is more than
one column section where the sum is nonzero then the total weight of the sum
is already at least 6.  Otherwise, the requirements of theorem 5b
are satisfied in the nonzero section.

\section{Outline of search procedure}

A search for $N_{1L}'$ configurations can be carried out inductively,
computing the isomorphism classes with $r$ rows from those with $r-1$ rows.
A bound can be imposed on the number of columns $c$ ($n-5$ in the preceding
section), since only configurations with large $r/c$ are of interest.
In this paper, the bound on $c$ is 18, and the maximum value of $r$
considered is 14 (although see section 4).

General procedures for isomorphism testing of incidence matrices exist,
including Brendan McKay's ``Nauty''
\cite{McKay}, and \cite{Leon}.  For one use of
Nauty in coding theory, see \cite{Jaffe}.  Discussions of using isomorphism
detection procedures in searching for combinatorial configurations can be
found in the literature, for example
\cite{Chen}, \cite{Kocay}, \cite{Margot}, and \cite{Raap}.
One common method is to convert each configuration as it is generated to a
``canonical representative'' of its equivalence class under isomorphism.

In this paper, specialized methods are used to reduce the matrix to one of
several ``partial canonicalizations''.  Each such is partitioned into 4 row
classes and 15 column classes.  Standard methods are then used to
canonicalize the partitioned partial canonicalizations, and the
lexicographically highest such is used.  Source code may be requested by
email from the author, and the description here omits various details.

An $N_1'$ configuration is assumed to be given as an incidence matrix $M$,
and a partition into 4 or fewer parts of the rows.
It may be assumed that the rows of a part are consecutive;
in some contexts missing parts are considered to be empty parts.
The parts are also called ``row classes''.

Numbering the parts from 0 to 3, a column may be given a type,
namely the function $f$ mapping $\{0,1,2,3\}$ to $\{0,1\}$,
where $f(i)=1$ iff the column has a one in part $i$.  The type
may be considered as a bit string of length 4, and denoted by a
hexadecimal digit 0-F (bit 0 being the low order bit).

Only configurations with no columns of type 0 need be considered.  Writing
the nonzero types in the order F7BDE3596AC1248, a configuration has a
``signature'', the 15-tuple of natural numbers which in each position gives
the number of columns of the corresponding type.

The symmetric group $S_4$ acts on the row classes, a permutation
$\alpha\in S_4$ being considered as``moving'' the class in position $i$ to
position $\alpha(i)$.  This induces an action on the column types, namely
$T\mapsto\alpha[T]$ where $T$ is the support
(note that, considering $T$ as a characteristic function,
$T'(p(i))=T(i)$ where $T'$ is the image).
Considering the signature to
be a function $\sigma$ from $\{T\}$ to the natural numbers, $\alpha$ acts
on the signatures by mapping $\sigma$ to $\sigma_{\alpha}$, where
$\sigma_{\alpha}(\alpha[T])=\sigma(T)$.

Given a signature $\sigma$, the ``canonicalized signature'' $\sigma^c$ is
defined to be the lexicographically greatest among the $\sigma_{\alpha}$.
As will be seen, it is useful to determine this, and also the right coset
$Gr$ of elements of $\alpha$ for which $\sigma_{\alpha}=\sigma^c$.
These can readily be determined by trying all 24 possiblities.  Indeed,
since signature canonicalization is of secondary cost, an efficient
implmentation of this method would undoubtedly suffice.  As will be seen,
one refinement was made.

For canonicalizing the signature it is useful to have a library of
routines for computing with permutations in $S_4$.  The permutations can
be ordered (a recursive order where the first 6 elements are $S_3$
was used), and tables coded which apply a permutation given as an
index in the order, by an array reference.

$S_4$ has 30 subgroups \cite{Maguit}.
Not all of them can occur as a stabilizer
of a signature, but it is simplest to code tables for all of them, in
particular a table of elements per subgroup index.  There are 234 cosets.
A  coset may be represented as a bit vector of length 24.  As noted above,
the right coset for a canonicalized signature is readily computed along
with it.  A hash table can be used to obtain the subgroup index and a
coset representative from the bit vector.

To speed up the process of canonicalizing the signature, the
columns may be grouped according to weight, and for each weight,
the canonicalized signature and right coset determined
successively.  For a given weight, only permutations in the
stabilizer of the higher weight columns need be considered.

For the weight 3 columns, the weights may be sorted.
There are 8 possibilities
$S_1RS_2RS_3RS_4$ where $R$ is $<$ or $=$ among the sorted sizes; each
yields a partition of $\{1,2,3,4\}$, and thereby a subgroup of $S_4$,
consisting of the product of the symmetric groups acting on the parts.
For the remaining weights, all possibilities within the stabilizer so far
are tried.  It might be possible to acheive a speed up in the case of weight
2 vectors with the full group acting, and this is certainly true in the case
of weight 1 vectors; but this was omitted.

The columns of type $f$ for some $f$ will be called a column class.
In addition to requiring an incidence matrix to have
the rows of each row class contiguous,
the columns of each column class will be required to be also.
Further, the column classes are required to be in the order given above.
A matrix $M$ with a such a row partition
consists of $60=4\times 15$ blocks, one for each row class and column class
(some of blocks may be empty, i.e., have 0 rows or 0 columns).

Supposing a method is specified for specifying the canonicalization $M^{cf}$
of an $N_1'$ configuration $M$ when the row classes are fixed, the
canonicalization of $M^c$ of $M$ may be defined as the lexicographically
highest of the $M_\alpha^{cf}$, where $M_\alpha$ are those matrices
obtained from $M$ by permuting the row classes, to yield the canonicalized
signature (i.e, where $\alpha\in Gr$ where $Gr$ is as above.

When applying a permutation of the row classes, the columns may be
permuted in any manner to obey the column restriction.
In obtaining $M^{cf}$, only permutations which preserve the blocks need be
considered.

At first, the author intended to write a canonicalization procedure
from scratch, under the belief that this would be faster and thus
more likey to complete.  However, Nauty has a reputation for being
fast; it permits specifying an initial partition, in this case into
blocks as above; and a preliminary version using it would permit debugging
the other code and provide a check.  A version using Nauty was thus
coded.

The generation of the configurations proceeds in stages, for $r$ increasing
up to some maximum value, where the number of columns is limited to some
maximum value.  At the beginning of the stage for $r$, the $r-1$ row
configurations are packed in an array; the rest of memory is used for a hash
table for the $r$ row configurations.  At the end of a stage, the hash table
is packed down to the beginning of memory.

Each $r-1$ row configuration is unpacked, and the span generated.
For each part of its row partition, a row is added in every possible way.
For each resulting configuration, a check is made whether it
is $N_{1L}$.  If so its signature is computed and canonicalized,
and $M^c$ is obtained as described above.  $M^c$ is added to the hash
table if it is not already in it.

\section{Results of search}

Initially the program was run with maximum values of 10 and 15 for $r$
and $c$.  This ran in 21 seconds, so the limits were raised to 12 and 18.
This run found that configurations with $r=12$ and $c=16$ exist.
The limit on $r$ was increased to 14.  The time for the run with
these values was 228 minutes.

For this paper, further increases to the limit were omitted, as this
would have required additional work.  For example,
the input graph to Nauty has 14+18=32 nodes.  If the graph has more than
32 nodes the rows of the adjacency matrix no longer fit in a word, and
Nauty's execution time would increase.  Again, though, see section 4.
For this paper, with the limits
of 14 and 18, the Nauty version is the final version.

Table 1 shows the number of isomprhism classes of $N_{1L}$
configurations with $r$ rows and $c$ columns, for $2\leq r\leq 14$ and
$5\leq c\leq 18$.

\begin{table*}
\begin{center}
\begin{tabular}{r|rrrrrrrrrrrrrr}
 $r$&   5&6&7&8&9&10&11&12&13&14&15&16&17&18\\
\hline
 2&   1&2&0&0&0&0&0&0&0&0&0&0&0&0\\
 3&   0&0&3&2&3&0&0&0&0&0&0&0&0&0\\
 4&   0&0&0&2&10&11&5&5&0&0&0&0&0&0\\
 5&   0&0&0&0&0&12&42&38&24&8&6&0&0&0\\
 6&   0&0&0&0&0&0&23&153&257&213&108&48&14&9\\
 7&   0&0&0&0&0&0&0&30&583&1635&1927&1262&607&223\\
 8&   0&0&0&0&0&0&0&5&13&2442&11813&18982&16261&9187\\
 9&   0&0&0&0&0&0&0&0&1&30&9153&87725&200690&219285\\
10&   0&0&0&0&0&0&0&0&0&1&170&26957&652926&2220665\\
11&   0&0&0&0&0&0&0&0&0&0&0&840&48624&4677339\\
12&   0&0&0&0&0&0&0&0&0&0&0&6&2513&85836\\
13&   0&0&0&0&0&0&0&0&0&0&0&0&24&3372\\
14&   0&0&0&0&0&0&0&0&0&0&0&0&0&100\\
\end{tabular}
\caption{Isomorphism class counts}
\end{center}
\end{table*}

From this, the value of $c_1$ is larger than 2/3.  Indeed, writing
$c_{\min}$ for the smallest $c$ for which configurations exist,
for even $r$ with $8\leq r\leq 14$, $c_{\min}$ increases by 2 as $r$ does.
This suggests that $c_1$ is at least 1.

In \cite{Dowd07} the following observations are made.
\begin{itemize}
\item An $N_{1L}'$ configuration with 2 flags in each column
is a cubic graph.
\item Such a cubic graph must be triangle free.
\item Two of the 6 cubic graphs on 8 vertices are triangle free.
\item Among the $N_{1L}'$ configurations with $r=8$ and $c=12$,
both triangle-free cubic graphs occur.
\end{itemize}
From the table, there are 5 configurations with $r=8$ and $c=12$.  It is
readily verified that 3 of these are the cube, and 2 are the other
possible cubic graph.  For all 5 configurations the list of partition
sizes is 2,2,2,2.

\section{A second search}

Since the ``gcc'' compiler for the ``x86'' processor supports a 64 bit
``long long'' type, a version of Nauty can be used where the rows of an
adjacency matrix fit in a ``long long''.  A partial search was conducted
using this feature.  Starting from the configurations with $r=13$, and
$c=17$ or $c=18$, only extensions by up to two columns were considered,
and only the minimum two $c$ values for each $r$ used as input to the next
stage.

This search yields an upper bound on $c_{\min}$.
The results are given in the following table.
After $r=19$ the program aborted due to insufficient memory.

\begin{table}[hb]
\begin{center}
\begin{tabular}{r|rrrrr}
$r$&15&16&17&18&19\\
\hline
bound&19&19&21&22&22\\
\end{tabular}
\end{center}
\end{table}

The results of this search suggest that the status of conjecture 2
is unclear.  The ratio $r/c$ increases, but so slowly that the computer
searches done here don't give any clear indication of its limiting
value.  For example, it is still open whether it can exceed 1.

It should also be noted that the best known linear double error correcting
codes do not contain $N_{1L}'$ configurations with values of $r/c$ as high
as those of configurations found here (see \cite{Dowd07}).

Topics for further research clearly include the following.
\begin{itemize}
\item An $N_{1L}$ configuration yields a linear double error correcting code
containing it, with $n=c+5$ and $k$ the rank of the configuration, augmented
with $v$.  This code may not be very good.  A ``goodness'' measure of
interest is $\left(1+n+{n\choose 2}\right)/2^(n-k)$.
\item Good codes for a given configuration, and configurations for a
given good code, should be more extensively investigated.
\item Methods for obtaining $N$ configurations from $N_1$ configurations
(for example using classical groups) might be of interest.
\item Although omitted here, determination of all configurations up to
$r=16$ and $c=19$ can be probably be achieved by the methods presented here.
\end{itemize}

\bibliography{n1lsarx}

\begin{thebibliography}{10}

\bibitem{BJL}
T.~Beth, D.~Jungnickel, and H.~Lenz.
\newblock {\em Design Theory}.
\newblock Cambridge University Press, 1993.

\bibitem{Chen}
L.~Chen.
\newblock Graph isomorphism and identification matrices: Parallel algorithms.
\newblock {\em IEEE Trans.\ Parallel Distrib.\ Syst.}, 7:308--319, 1996.

\bibitem{Dowd88}
M.~Dowd.
\newblock Questions related to the {E}rdos-{T}uran conjecture.
\newblock {\em SIAM Journal in Discrete Mathematics}, 1:142--150, 1988.

\bibitem{Dowd07}
M.~Dowd.
\newblock Configurations in binary linear codes.
\newblock In {\em Fourth International Conference on Applied Mathematics and
  Computing}, 2007.
\newblock Full text available at www.hyperonsoft.com.

\bibitem{Jaffe}
D.~Jaffe.
\newblock Binary linear codes: New results on nonexistence.
\newblock draft, www.math.unl.edu/\lower.5ex\hbox{\~{}}djaffe2, 2000.

\bibitem{Kaski}
P.~Kaski.
\newblock Algorithms for classification of combinatorial objects.
\newblock Technical Report~94, Helsinki University of Technology Laboratory for
  Theoretical Computer Science, 2005.

\bibitem{Kocay}
W.~Kocay.
\newblock On writing isomorphism programs.
\newblock In W.~D. Wallis, editor, {\em Computational and Constructive Design
  Theory}. Kluwer, 1996.

\bibitem{Leon}
J.~Leon.
\newblock {\em Partition Backtrack Programs: User's Manual}.
\newblock www-groups.dcs.st-and.ac.uk/\lower.5ex\hbox{\~{}}gap/\%Manuals/
  pkg/guava2.7/src/leon/doc/leon\_guava\_manual.

\bibitem{Maguit}
A.~Maguitman.
\newblock S4 visualized.
\newblock www.cs.indiana.edu/\lower.5ex\hbox{\~{}}anmaguit/s4gallery.

\bibitem{Margot}
F.~Margot.
\newblock Pruning by isomorphism in branch-and-cut.
\newblock {\em Mathematical Programming}, 94:71--90, 2002.

\bibitem{McKay}
B.~McKay.
\newblock Practical graph isomorphism.
\newblock {\em Congressus Numerantium}, 30:45--87, 1981.

\bibitem{Raap}
S.~Raaphorst.
\newblock Branch-and-cut for symmetrical {ILPs} and combinatorial designs.
\newblock Master's thesis, School of Information Technology and Engineering,
  University of Ottawa, 2004.

\end{thebibliography}

\end{document}